\documentclass[sn-nature]{sn-jnl}

\usepackage{graphicx}%
\usepackage{multirow}%
\usepackage{amsmath,amssymb,amsfonts}%
\usepackage{amsthm}%
\usepackage{mathrsfs}%
\usepackage[title]{appendix}%
\usepackage{xcolor}%
\usepackage{textcomp}%
\usepackage{manyfoot}%
\usepackage{booktabs}%
\usepackage{algorithm}%
\usepackage{algorithmicx}%
\usepackage{algpseudocode}%
\usepackage{listings}%

\usepackage{crimson}
\usepackage[switch]{lineno}

\theoremstyle{thmstyleone}%
\theoremstyle{thmstyletwo}%

\theoremstyle{thmstylethree}%

\begin{document}

\title[Article Title]{False Consensus Biases AI Against Vulnerable Stakeholders}


\author*[1]{\fnm{Mengchen} \sur{Dong}}\email{dong@mpib-berlin.mpg.de}

\author[2]{\fnm{Jean-Fran\c{c}ois} \sur{Bonnefon}}\email{jean-francois.bonnefon@tse-fr.eu}

\author[1]{\fnm{Iyad} \sur{Rahwan}}\email{rahwan@mpib-berlin.mpg.de}

\affil*[1]{\orgdiv{Center for Humans and Machines}, \orgname{Max Planck Institute for Human Development}, \orgaddress{\street{Lentzeallee 94}, \city{Berlin}, \postcode{14195}, \country{Germany}}}

\affil[2]{\orgdiv{Toulouse School of Economics}, \orgname{University of Toulouse 1 Capitole}, \orgaddress{\street{1 Esp. de l'Université}, \city{Toulouse}, \postcode{31000}, \country{France}}}


\abstract{The deployment of AI systems for welfare benefit allocation allows for accelerated decision-making and faster provision of critical help, but has already led to an increase in unfair benefit denials and false fraud accusations. Collecting data in the US and the UK ($N = 2\ 449$), we explore the public acceptability of such speed-accuracy trade-offs in populations of claimants and non-claimants. We observe a general willingness to trade off speed gains for modest accuracy losses, but this aggregate view masks notable divergences between claimants and non-claimants. Although welfare claimants comprise a relatively small proportion of the general population (e.g., 20\% in the US representative sample), this vulnerable group is much less willing to accept AI deployed in welfare systems, raising concerns that solely using aggregate data for calibration could lead to policies misaligned with stakeholder preferences. Our study further uncovers asymmetric insights between claimants and non-claimants. The latter consistently overestimate claimants’ willingness to accept speed-accuracy trade-offs, even when financially incentivized for accurate perspective-taking. This suggests that policy decisions influenced by the dominant voice of non-claimants, however well-intentioned, may neglect the actual preferences of those directly affected by welfare AI systems. Our findings underline the need for stakeholder engagement and transparent communication in the design and deployment of these systems, particularly in contexts marked by power imbalances.
\newline
\newline
\newpage
\textbf{Significance Statement. }Artificial Intelligence should optimally share the values and goals of humans. But which humans should AI align with, given its potential for social good? In the context of social welfare distribution, we analyze public preferences for the trade-offs between speed gains and accuracy losses when AI systems are deployed. We show that welfare claimants can be more averse to AI deployed in welfare systems, and less vulnerable non-claimants can misunderstand their preferences for the trade-offs. These results constitute a strong call for engaging specifically with vulnerable stakeholders when designing government AI systems, rather than relying on aggregate data or assuming that other stakeholders with the dominant voice in society can understand their preferences.}




\maketitle

\section{Introduction}\label{sec1}

The use of Artificial Intelligence (AI) is becoming commonplace in government operations \cite{misuraca2020ai, engstrom2020government, coglianese2020ai, desousa2019how}. In the United States alone, a 2020 survey of 142 federal agencies found that 45\% were using or planning to use machine learning algorithms to streamline their operations, increase their capacities, or improve the delivery of their public services \cite{engstrom2020government}. In the specific context of providing welfare benefits, the main promise of AI is to speed up decisions \cite{bansak2018improving,desousa2019how}. For many individuals and families, welfare benefits provide critical assistance in times of financial hardship or emergency. Using AI to speed up decisions can avoid delays that would exacerbate these hardships, and decrease the period of uncertainty and anxiety during which applicants are waiting for a decision. There is, however, a documented risk that since welfare AI systems often focus on fraud detection, their speed gains come with a biased accuracy loss, increasing the acceptable trade-offs between the speed and accuracy of welfare allocations rate at which people are unfairly denied the welfare benefits they are entitled to \cite{alston2015report,booth2019benefits,constantaras2023inside,muralidharan2020identity,mosley2023algorithm, carney2020artificial}.

As a result, government agencies that seek to deploy welfare AI systems must strike a careful balance between speed gains and accuracy losses, and this balance must be informed by public preferences \cite{noriega2018algorithmic, bonnefon2020moral}, for at least two reasons. First, we know that people who lose trust in the AI used by one government agency also lose trust in the AI used by other  government agencies---if welfare AI systems ignore public preferences when balancing speed and accuracy, they risk creating distrust that can bleed into perceptions of other government services \cite{longoni2023algorithmic, dietvorst2015algorithm}. Second, and more immediately, the wrong balance of speed gains and accuracy losses could erode the trust of people who need welfare benefits, and make them less likely to apply, for fear of being wrongly accused of fraudulent claims \cite{longoni2023algorithmic}, especially when the AI system is labeled with foreboding names like `FraudCaster' \cite{Pondera2022FraudCaster} or described as a `suspicion machine' in the media \cite{arstechnica2022algorithms, constantaras2023inside}. In sum, it is important for welfare AI systems to trade off speed and accuracy in a way that is aligned with the preferences of the general public as well as with the preferences of potential claimants.

Great efforts have been made to understand people’s attitudes toward and concerns about welfare AI systems, often focusing on the opinions of the general public \cite{longoni2023algorithmic, jonsson2021european} or vulnerable populations directly affected by welfare AI systems \cite{carney2020artificial, brown2019toward}. Qualitative evidence has also been accumulated regarding the divergent preferences of different stakeholders involved in AI governing systems \cite{constantaras2023inside, lee2017human}, contributing to long-lasting philosophical and regulatory discussions on fairness and alignment principles \cite{arnstein1969ladder, birhane2022power, reisman2018algorithmic, huang2019veil}. However, less is known about the extent of divergence in AI design preferences and reconciliation between different perspectives and interests.

Here we present experimental evidence on two critical challenges for aligning AI with human values in welfare AI systems. First, we identify heterogeneous preferences of welfare claimants versus non-claimants, with claimants showing a stronger AI aversion irrespective of AI trades off speed and accuracy. Second, we show that while welfare claimants show insights into the preferences of non-claimants, non-claimants show no insights into the preferences of claimants. In other words, the perspective of non-claimants is relatively easy to understand, but only claimants understand their own perspective. These results hold in a representative US sample and in a targeted UK sample balancing the number of claimants and non-claimants. The combination of heterogeneous preferences and asymmetric insights creates the risk that welfare AI systems be aligned with the position of the largest, best understood, least vulnerable group – silencing the voice of the smallest, least understood, most vulnerable group, which nevertheless comprises the primary stakeholders in the deployment of welfare AI.

\begin{figure}
\includegraphics[width = \textwidth]{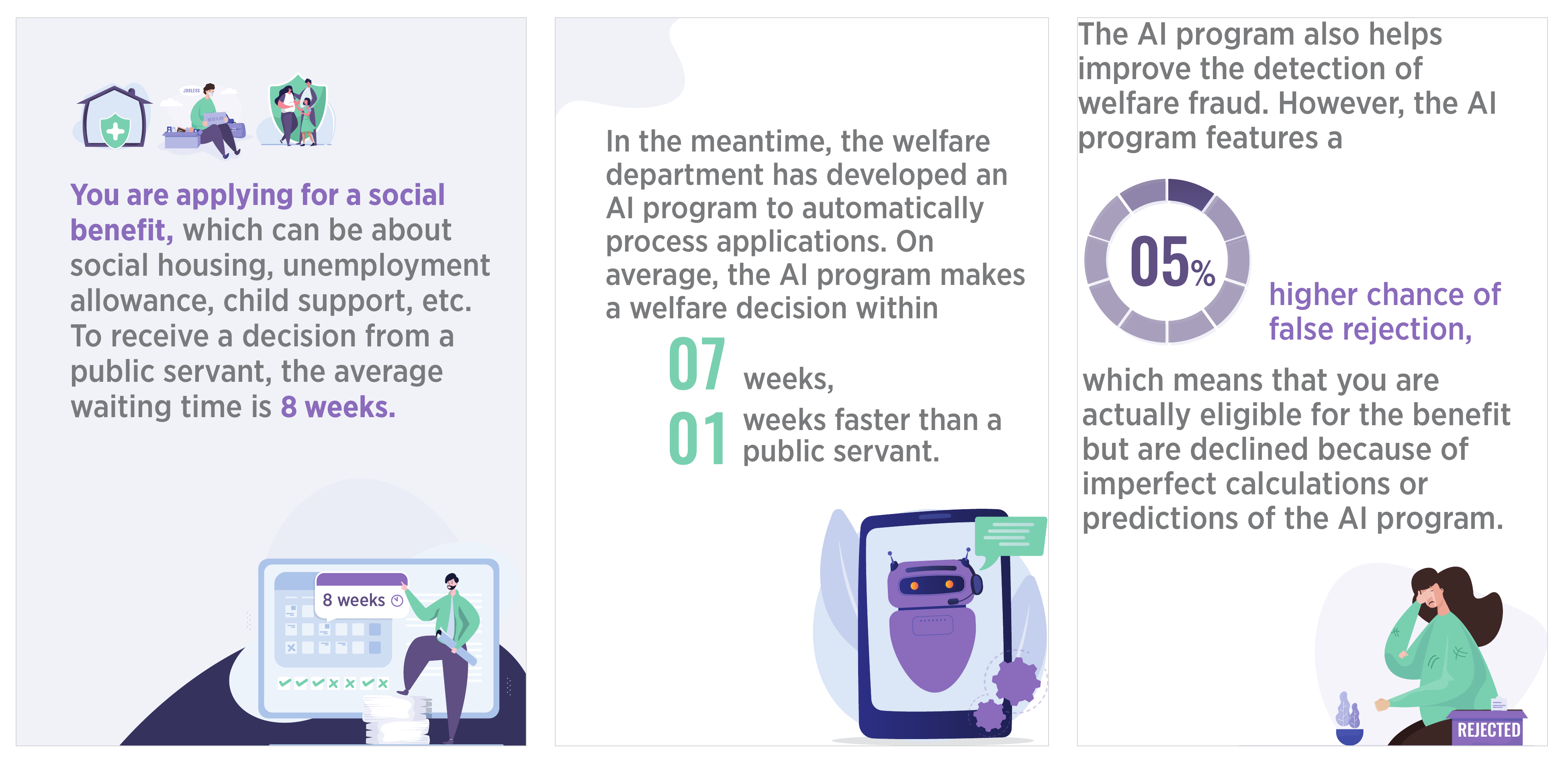}
\caption{\emph{An example of experimental stimuli, where the AI system is one week faster than humans but leads to a 5\% accuracy loss. The complete list of stimuli consisted of 36 such trade-offs, combining speed gains of 1 to 6 weeks (by the increment of 1) and accuracy losses from 5\% to 30\% (by the increment of 5\%). \label{fig:tradeoff}}}
\end{figure}

\section{Results in the US}\label{sec2}

Participants in the US ($N = 987$, representative on age, gender, and ethnicity, 20\% self-declaring as welfare claimants) indicated their preference between human and AI welfare decisions. We varied the information about speed gains (1/2/3/4/5/6 weeks faster, as compared to a baseline waiting time of 8 weeks if handled by public servants) and accuracy losses (5/10/15/20/25/30\% more false rejections than public servants) within a realistic range, based on governmental reports and third-party investigations \cite{muralidharan2020identity,noriega2018algorithmic, plc2021annual, benedikt2020human}, yielding 36 trade-offs (as illustrated in Figure~\ref{fig:tradeoff}). In each trade-off condition, participants indicated their preference on a scale ranging from 0 = \emph{definitely a public servant} to 100 = \emph{definitely the AI program}. Participants were randomly assigned to respond from their own perspective as claimants or non-claimants, or to adopt the opposite perspective. The US study was not preregistered, hence all analyses should be considered exploratory.

\begin{figure}[H]
\centering
\includegraphics[width = 0.6\textwidth]{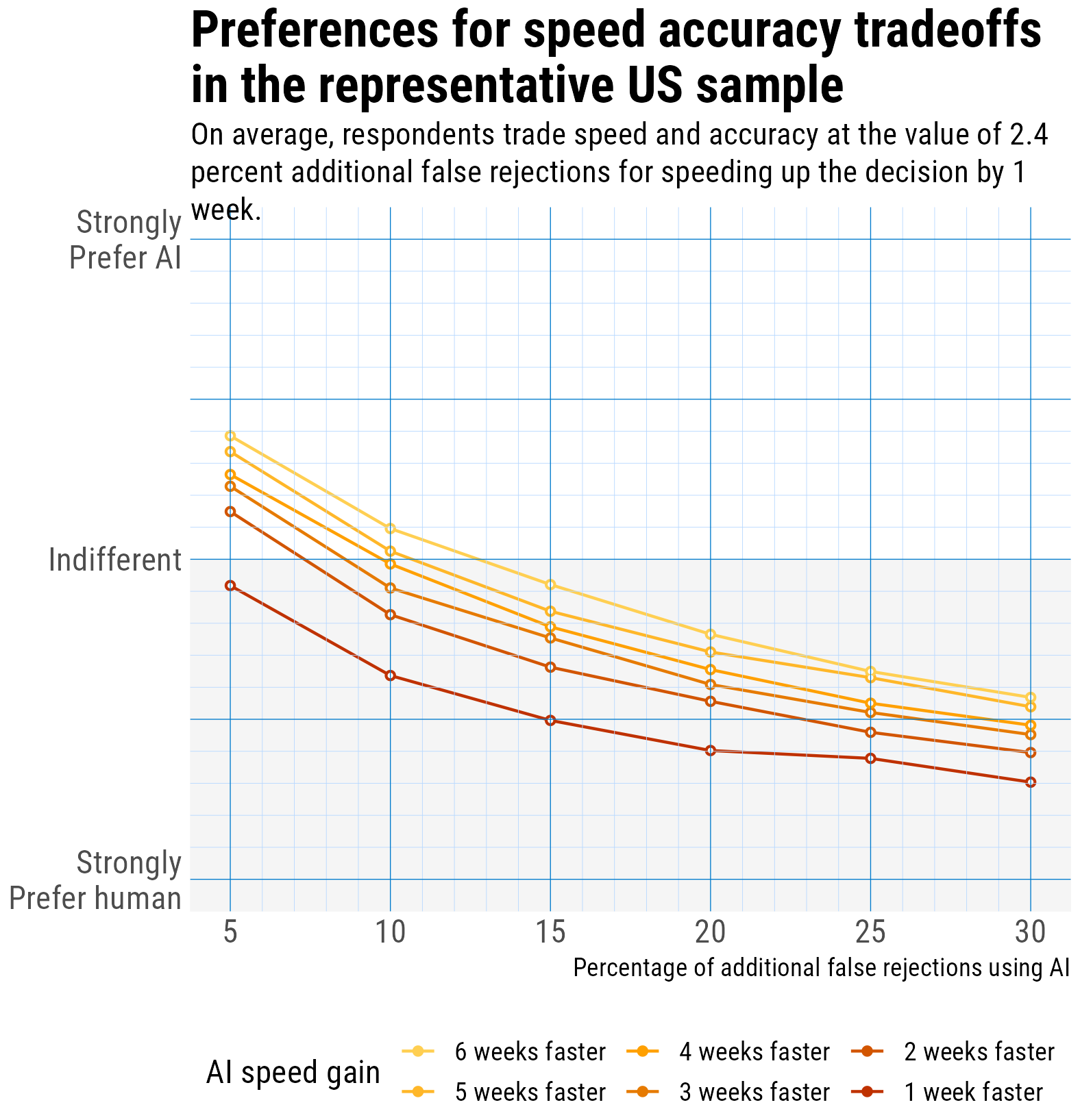}
\caption{\emph{Preferences for speed accuracy trade-offs from own perspective, in the representative US sample ($N = 506$; 21\% as welfare claimants).}\label{fig:study1groups}}
\end{figure}

When participants responded from their own perspective ($N = 506$), their willingness 
to let AI make decisions was influenced both by speed gains ($\beta = 0.19$, $p < .001$) and accuracy losses ($\beta = 0.40$, $p < .001$). Overall (see Figure \ref{fig:study1groups}), they traded off a 1-week speed gain for a 2.4 percentage points loss of accuracy. Among these US participants, 21\% self-declared as welfare claimants. For all the 36 trade-offs, these claimants (vs. non-claimants) showed greater average aversion to letting AI make welfare decisions ($\beta = -0.19$ $p < .001$). The average difference between the responses of claimants and non-claimants was 5.9 points (range: 0.3 to 12.8, see Figure \ref{}A). 

Figure \ref{}B displays the biases of claimants and non-claimants when trying to predict the answers of the other group, across the 36 tradeoffs. Here we calculate the bias for each trade-off condition by subtracting participants’ actual preference (e.g., claimants taking a claimant perspective) from the other groups’ insights through perspective taking (e.g., non-claimants taking a claimant perspective). We then compare the bias scores with zero to determine their statistical significance, using the formula below:

\[bias_{ij} = \beta_0 + \mu_{0j}  + \epsilon_{ij}  \]
where $bias_{ij}$ represents the bias for the $i$th observation in the jth participant, $\beta_0$  represents the fixed intercept, $\mu_{0j}$ represents the random effect for the $j$th participant, and $\epsilon_{ij}$ represents the residual error for the $i$th observation in the $j$th participant.

Both groups fail to completely take the perspective of the other group. On average, claimants underestimate the answers of non-claimants by 4.8 points, and non-claimants overestimate the answers of claimants by 6.4 points. Both biases are significantly different from zero ($p = .032$ for claimants, and $p < .001$ for non-claimants): the 95\% confidence interval is [-9.2, -0.5] for claimants, and [4.4, 8.4] for non-claimants. Two issue when comparing the biases between the two groups, though, are their unequal size in our sample (the standard error for claimants is twice that for non-claimants), and the lack of financial incentives for responding correctly. These two issues are addressed in our second study.

In sum, data from our representative US sample shows that US citizens, on average, were willing to trade a 2.4 accuracy loss for a 1-week speed gain. However, welfare claimants are systematically more averse to AI than non-claimants, and we find evidence for a small asymmetry in the insights that claimants and non-claimants have into each other's answers, with claimants being more calibrated when predicting the answers of non-claimants.

\begin{figure}
\includegraphics[width = \textwidth]{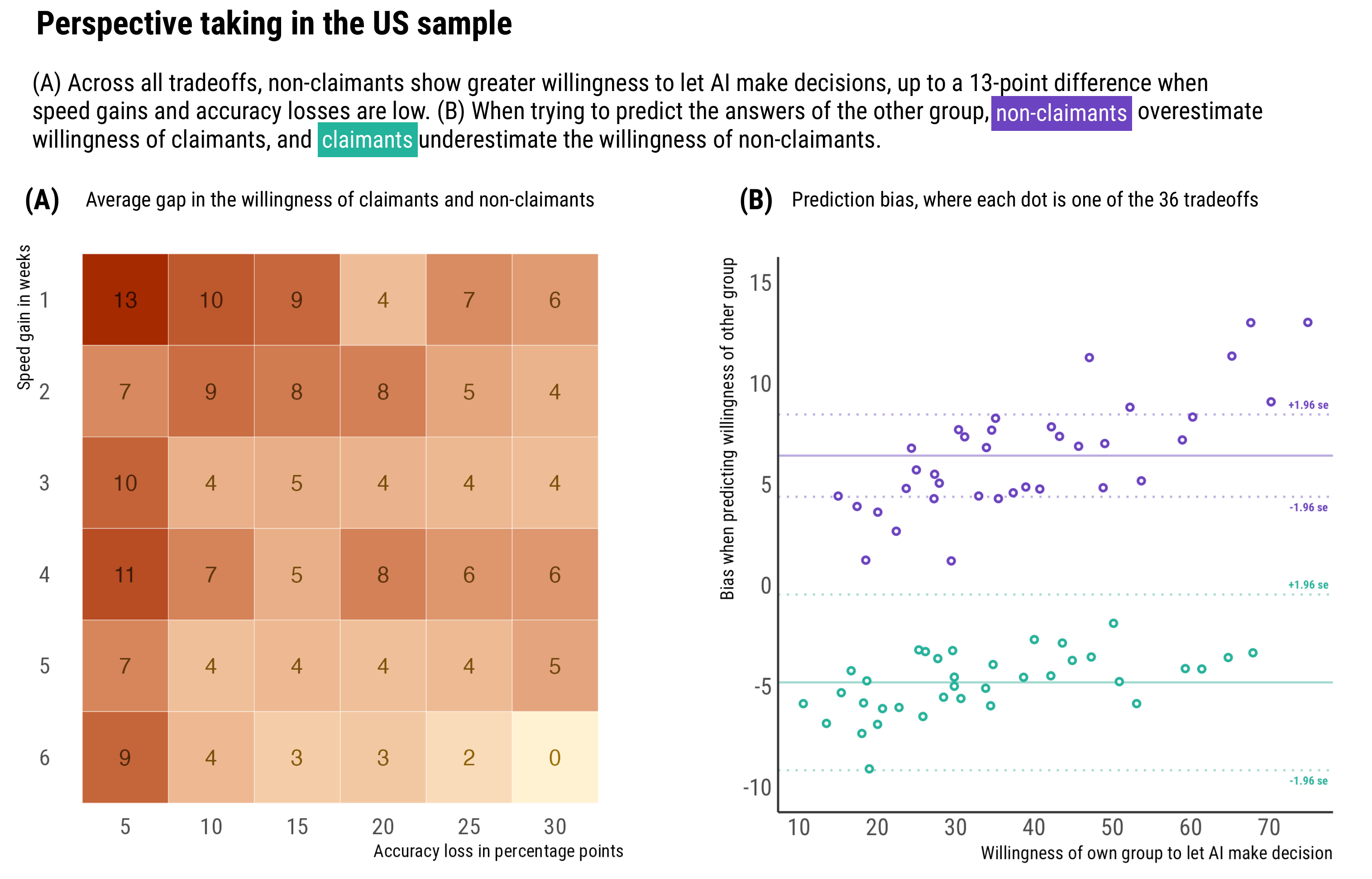}
\caption{\emph{Perspective taking in the representative US sample (N = 987; 20\% as welfare claimants). (A) The average gap between the willingness of claimants and non-claimants to let AI make welfare decisions across the 36 tradeoffs. (B) Biases of claimants and non-claimants trying to predict the answers of the other group.}\label{}}
\end{figure}

\section{Results in the UK}\label{sec3}

To replicate and extend the results obtained from the US representative sample, we collected data from $N = 1\ 462$ participants in the UK with a balanced composition of claimants and non-claimants. Such a balanced sample can help consolidate our pre-registered test on the asymmetry in perspective-taking. In addition, we implemented the following changes: 

\begin{enumerate}
 \item We examined preferences about a specific benefit in the UK (the Universal Credit) and targeted a balanced sample between Universal Credit claimants (48\%) and non-claimants (52\%). The UK government has recently announced the deployment of AI for the attribution of this benefit, raising concerns that the AI system may be biased against some claimants \cite{booth2019benefits}. 
 \item We adopted a different range of speed (0/1/2/3 weeks faster, as compared to a baseline waiting time of 4 weeks if handled by public servants) and accuracy (0/5/10/15/20\% more false rejections than public servants) parameters, resulting in 20 trade-offs. Notably, when welfare AI demonstrates comparable performance (i.e., 0 week faster and 0\% more error), people were still in favor of humans making welfare decisions ($M = 45.4$, $SD = 28.7$; $t = 4.36$, $p < .001$). 
\item We added financial incentives for participants to correctly predict the preferences of the other group, that is, when non-claimants predict claimants’ preference and claimants predict non-claimants’ preference. We also asked non-claimants whether they had claimed welfare benefits in the past, whether they thought they may claim benefits in the future, and whether they were acquainted with people who were welfare claimants, to assess whether these circumstances made it easier to adopt the perspective of claimants.
\item For each trade-off, we additionally asked participants whether their trust in the government would decrease or increase (from 0 = \emph{decrease a lot} decrease a lot to 100 = \emph{increase a lot}) if the government decided to replace public servants with the AI program they just considered.
\item Finally, we added a treatment that made explicit the existence of a procedure to ask for redress in case a claimant felt their claim was unfairly rejected. Even though participants in the human redress condition believed in the chance to appeal ($\beta = 0.37$, $p < .001$; vs. the redress condition), this clarification did not impact trade-off preferences ($\beta = 0.03$, $p = .210$). Therefore, we pool the data from this treatment with that of the baseline treatment. Full analyses of this treatment are presented in the Supplementary Information.
\end{enumerate}

\begin{figure}
\centering
\includegraphics[width = 0.6\textwidth]{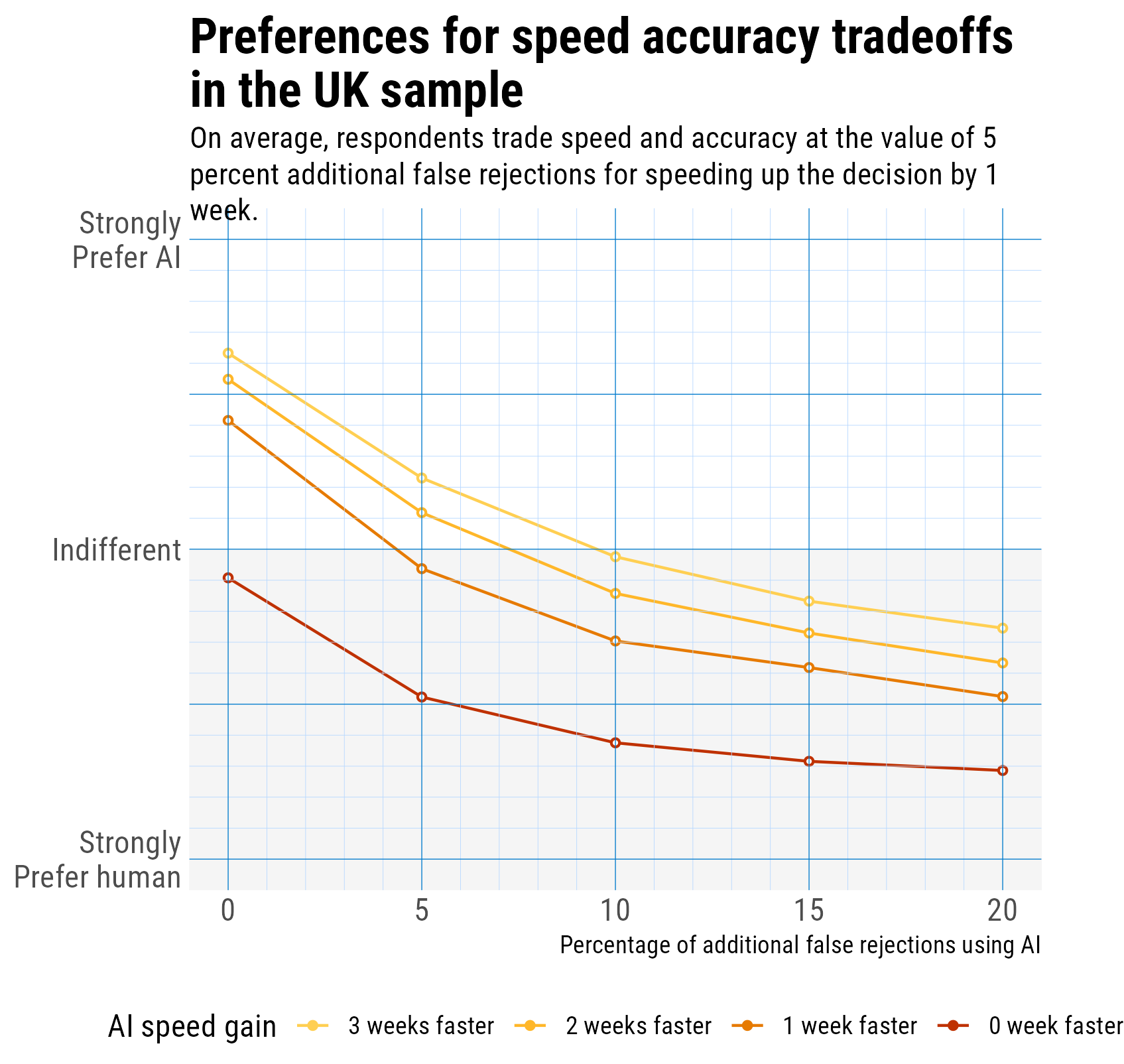}
\caption{\emph{Preferences for speed accuracy trade-offs from own perspective, in the balanced UK sample (N = 739; 47\% as welfare claimants)}\label{fig:study2groups}}
\end{figure}

Again, when participants responded from their own perspective ($N = 739$), their willingness to let AI make decisions was influenced both by speed gains ($\beta = 0.34$, $p < .001$) and accuracy losses ($\beta = 0.44$, $p < .001$). Overall (see Figure~\ref{fig:study2groups}), they traded off a 1-week speed gain for a 5 percentage point loss of accuracy.  Among these UK participants, 47\% self-declared as current claimants of the Universal Credit. As in the US study, for all 20 trade-offs, welfare claimants showed greater average aversion to letting AI make welfare decisions ($\beta = -0.09$, $p = .008$), with an average difference of 5.7 points (range: 0.1 to 8.7, see Fig. 5A). In both groups, we observe a strong correlation across trade-offs between the aversion to letting the AI make decisions, and the loss of trust in the government that would deploy this AI (r = .77 for claimants, and r = .84 for non-claimants). 

\begin{figure}
\includegraphics[width = \textwidth]{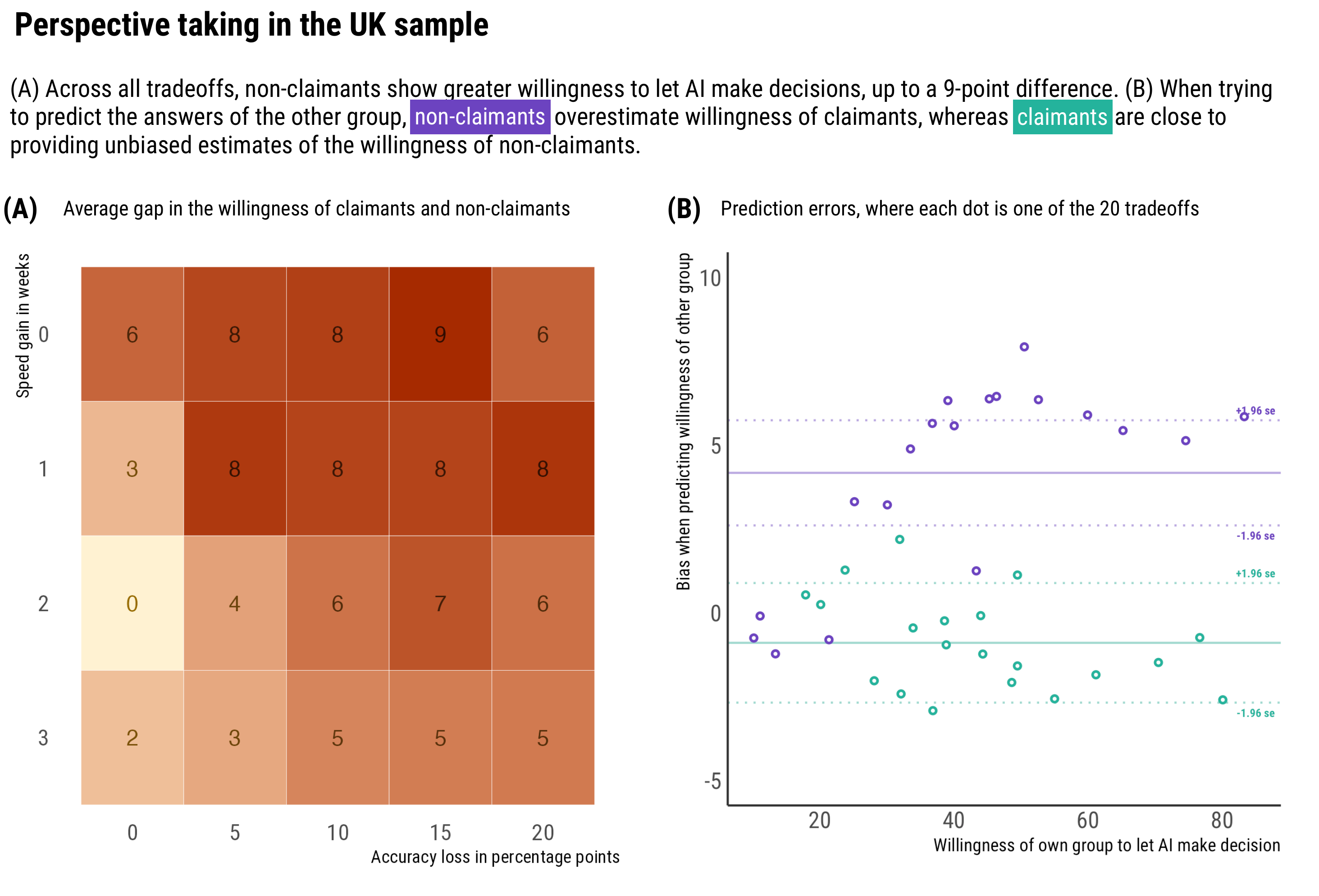}
\caption{\emph{Perspective taking in the balanced UK sample (N = 1462; 48\% as welfare claimants). (A) The average gap between the willingness of claimants and non-claimants to let AI make welfare decisions across the 20 tradeoffs. (B) Biases of claimants and non-claimants trying to predict the answers of the other group.}\label{fig:study2insight}}
\end{figure}

Figure \ref{fig:study2insight}B displays the biases of claimants and non-claimants when trying to predict the answers of the other group, across the 20 trade-offs. As in the US study, we calculated the perspective-taking biases for claimants and non-claimants, respectively. On average, claimants provide an unbiased estimate of the answers of non-claimants ($p = .323$), with an underestimation of 0.9 points and a 95\%-confidence interval including zero, [-2.7, 0.9]. Non-claimants, however, overestimate the preferences of claimants by 4.2 points ($p < .001$), with a 95\%-confidence interval of [2.6, 5.7]. These asymmetrical insights between claimants and non-claimants are consistent with our preregistered prediction. To explore whether some life experiences may reduce bias in the predictions of non-claimants, we recorded whether they had past experience as claimants of other benefits, whether they were acquainted with current claimants, and their perceived likelihood of becoming claimants in the near future. We found no credible evidence for any of these effects.

In sum, results from our targeted UK sample consolidate and extend results from our representative sample. The average willingness to trade a 5-point accuracy loss for a 1-week speed gain hides heterogeneity in responses, with welfare claimants being systematically more averse to AI than non-claimants. We also find strong evidence for asymmetrical insights between claimants and non-claimants: claimants are well-calibrated when predicting the answers of non-claimants, but non-claimants overestimate the willingness of claimants to let AI make decisions. Finally, lower acceptance of the AI system for welfare allocation is strongly linked to decreased trust in the government among both welfare claimants and non-claimants.

\section{Discussion}\label{sec4}

One primary advantage of using AI for welfare benefit allocation is quicker decision-making, allowing claimants to receive support faster \cite{desousa2019how, bansak2018improving}. However, these systems often result in an accuracy loss, potentially leading to unfair denials or false fraud accusations \cite{alston2015report, booth2019benefits, constantaras2023inside, muralidharan2020identity, mosley2023algorithm, carney2020artificial}. Governments must carefully balance these trade-offs to maintain public trust \cite{longoni2023algorithmic, fiske2014gaining}. Indeed, we found that the acceptability of this balance to participants was closely tied to their resulting trust in the government.

In both the US and UK, our data suggested that participants would trade a one-week speed gain for a 2.5 to 5 percentage point accuracy loss. However, we also found that averaging across participants masked strong divergences between claimants and non-claimants. Though the difference between the two groups varied across trade-offs, welfare claimants were systematically less amenable to AI deployment than non-claimants. In parallel with the comparison between welfare claimants and non-claimants, we also conducted latent profile analysis to explore the underlying patterns in the preference data without relying on existing labels (e.g., claimant status). Since we did not find strong support for a particular number of profiles, we report the results in the Supplementary Information. In summary, using aggregate data to calibrate AI in welfare systems could backfire, as average responses may not reflect the divergent preferences of stakeholders. This finding aligns with recent calls in behavioral science to focus on heterogeneity when informing policy \cite{bryan2021behavioural}, as well as to consider the positionality of AI models \cite{cambo2022model}, that is, their social and cultural position with regard to the stakeholders with which they interface.

Data revealed a further complication: asymmetric insight between claimants and non-claimants. Claimants could provide unbiased estimates of the preferences of non-claimants, but non-claimants failed to do the same, even in the presence of financial incentives. These findings echo laboratory results suggesting that participants who are or feel more powerful struggle to take the cognitive perspective of others \cite{blader2016looking,galinsky2006power,guinote2017power,brown2017relationship}, as well as sociological theories positing that marginalized groups have greater opportunities and motivations to develop an understanding of the thoughts and norms of dominant groups \cite{Bourdieu1984,du2008souls,Haraway1988}.

In the context of welfare AI, asymmetric insights create the risk that the perspective of claimants may be silenced even when non-claimants seek to defend the interests of claimants. These well-intentioned non-claimants may use their dominant voice to shape public opinion and policy without realizing that they do not in fact understand the preferences of claimants, resulting in AI systems that are misaligned with the preferences of their primary stakeholders. Our results thus underline the need to actively engage with claimants when building welfare AI systems, rather than to assume that their preferences are well-understood or can be understood through empathetic perspective-taking. 

Our results also highlight the need for transparent communication about welfare AI systems’ design choices. Given the significant heterogeneity in public preferences, and the close link between meeting these preferences and trust in government, clear justifications are crucial. Despite increasing technical attempts to align AI with pluralistic values and diverse perspectives \cite{sorensen2024roadmap, bakker2022fine, conitzer2024social}, there are inevitably situations where agreement or reconciliation cannot be easily achieved (e.g., when non-claimants fail to estimate welfare claimants’ aversion to AI, but not vice versa). In this case, our study sheds light on the possibility of evidence-based public communication when human-centered AI designs need to de-emphasize the preferences of the general population but optimize toward a particular subgroup of people. Our core findings, heterogeneity and asymmetric insights, may also hold in other cases where AI is deployed in a context of power imbalance – conducting behavioral research on these cases in advance of AI deployment may help avoid the scandals that marred the deployment of welfare AI.

\section{Methods}\label{sec5}

Both of the US and UK studies were approved by the ethics committee at the Max Planck Institute for Human Development, and obtained informed consent from all participants. Data were collected in February 2022 and September 2022, respectively. All participants were recruited on Prolific for a study named “Artificial Intelligence in Social Welfare”, and were paid £1.6 upon completion. Participants in the UK study who had to predict the answers of the other group received an additional £0.03 for each response that fell within 5 points of this other group's average.

Both studies were hosted on Qualtrics. After providing informed consent and basic demographic information (age, gender, education, income, and political ideology), participants were instructed to take a claimant or non-claimant perspective. To familiarize themselves with the stimuli and response scale, they were first shown two extreme trade-offs in the survey, as training examples. They answered these two examples, and had a chance to review and change their answers. Then the survey started, and all targeted trade-offs were shown in random order, not including the two trade-offs that were shown as examples during the training phase. Complete descriptions of our design materials, and survey questions are included in the Supplementary Information. 

\subsection{The US study}\label{subsec1}

\bmhead{Participants} We had $N = 987$ participants from the United States, who were representative on age ($M = 45.3$, $SD = 16.3$), gender (473 males and 514 females), and ethnicity (77.8\% White, 11.4\% Black, 6.1\% Asian, 2.5\% Mixed, and 2.1\% other), and 20.4\% of them self-reported as welfare claimants at the time of the study. The sample size was determined based on the recent recommendation of around 500 people for latent profile analysis \cite{spurk2020latent}. We aimed for an almost doubled sample size given our two-condition perspective-taking manipulation. 

\bmhead{Design and procedure} The US study employed a mixed design, with one between-subjects and two within-subjects factors. First, participants were asked for their basic demographic information, and were randomly assigned to take either a claimant (“You are applying for a social benefit”) or a controlled taxpayer (“Someone else in your city is applying for a social benefit”) perspective. We then manipulated the information about welfare AI’s speed (6 conditions: 1/2/3/4/5/6 weeks faster than a public servant) and accuracy (6 conditions: 5/10/15/20/25/30\% more false rejections than a public servant). The presented speed (an average of 8 weeks) and accuracy (at most 40\% errors) baselines referred to realistic information from some governmental reports and third-party investigations \cite{muralidharan2020identity,noriega2018algorithmic, plc2021annual, benedikt2020human}. 

After knowing the perspective they should take, participants went through two training examples, reading two extreme cases of welfare AI (bad case: 0 week faster + 50\% more false rejections; good case: 7 weeks faster + 1\% more false rejections) and answering the same question “To what extent do you prefer a public servant or the AI program to handle your/the person’s welfare application?” on a 100-point scale (from 0 = \emph{definitely a public servant} to 100 = \emph{definitely the AI program}). They then had a chance to review and calibrate their answers in the two cases before moving to the 36 official test rounds – which did not allow revisions anymore. In each of the 36 test rounds, they read information about their perspective, AI speed, and AI accuracy in three consecutive cards (see Fig. S1 in the Supplementary Information for an illustration of the cards in different experiment conditions). After reading the three cards in each round, participants answered the same question about their preference for welfare AI versus public servants.  

\subsection{The UK study}\label{subsec2}

\bmhead{Participants} We performed a simulation-based power analysis for multilevel regression models, which suggested that a sample of $N = 800$ would allow us to detect the interaction effect of AI performance, claimant status, and perspective-taking with higher than 80\% power at an alpha level of 0.05 (see the pre-registration at \url{https://tinyurl.com/welfareAIregistration}). We therefore aimed for $N = 1600$ participants in the United Kingdom given our additional between-subjects human redress manipulation. As pre-registered, we filtered out participants who provided different answers to one identical welfare status question (“Are you a recipient of Universal Credit?”; Answer: “Yes/No”), which was embedded both in the Prolific system screener and our own survey. After the screening, we eventually had $N = 1\ 462$ participants (age: $M = 37.6$, $SD = 11.1$; ethnicity: 88.4\% White, 3.0\% Black, 5.6\% Asian, 2.7\% Mixed, and 0.3\% other), with relatively balanced compositions of males and females (42.7\% male, 55.9\% female, 1.4\% other), and welfare claimants (47.9\%) versus non-claimants (52.1\%). 

\bmhead{Design and procedure}  Study 2 examined a real-life social benefits scheme in the UK – Universal Credit (\url{https://www.gov.uk/universal-credit}). We employed a mixed design with three between-subjects and two within-subjects factors. As between-subjects factors, we recruited both Universal Credit claimants and non-claimants, and randomly assigned them to take a Universal Credit claimant or a controlled taxpayer perspective. They were then randomly assigned to a no redress or a human redress condition, which differed on whether claimants could appeal to public servants. As within-subject factors, we manipulated information about welfare AI’s speed (0/1/2/3 weeks faster, as compared to a baseline waiting time of 4 weeks if handled by public servants), and accuracy (0/5/10/15/20\% more false rejection). 

Before starting the 20 rounds of official tradeoff evaluations, as in the US study, participants went through two training examples with a chance of revision, reading two extreme cases of welfare AI (bad case: 0 week faster + 40\% more false rejections; good case: 3 weeks faster + 1\% more false rejections). In each example, they answered two questions on a 100-point scale: “To what extent do you prefer a public servant or the AI program to handle your/the person’s welfare application?” (0 = \emph{definitely a public servant} to 100 = \emph{definitely the AI program}) and “If the UK government decided to replace some public servants with the AI program in handling welfare applications, would your trust in the government decrease or increase?” (0 = \emph{decrease a lot} to 100 = \emph{increase a lot}). They then had a chance to review and change their answers in the two cases before moving to the 20 official test rounds – which did not allow revisions anymore. In each of the 20 test rounds, they read information about their perspective, AI speed, AI accuracy, and human redress condition in three consecutive cards (see Fig. S2 in the Supplementary Information for an illustration of the cards in different experiment conditions). After reading the three cards in each round, participants answered the same two question about their preference for welfare AI versus public servants, and their trust in the government. 

To increase the motivation of perspective taking, participants were informed and incentivized to take the opposite perspective, for each accurate answer that fell within ±5 points of the other group’s average. At the end of the 20 official rounds, as a manipulation check, participants indicated the extent to which they believed that “you/the person can appeal to public servants if you/they are not satisfied with the welfare decision made by the AI program?” (0 = \emph{not at all} to 100 = \emph{very much}). 

\subsection{Data availability}\label{subsec3}

All anonymized data can be found on Open Science Framework (at \url{https://tinyurl.com/welfareAI}; and will be made publicly accessible upon acceptance of the work). 

\subsection{Code availability}\label{subsec4}
All code necessary to reproduce all analyses can be found on Open Science Framework (at \url{https://tinyurl.com/welfareAI}; and will be made publicly accessible upon acceptance of the work). 

\section*{End notes}

\begin{itemize}
\item Acknowledgements: Agence Nationale De La Recherche ANR-19-PI3A-0004 (JFB); Agence Nationale De La Recherche ANR-17-EURE-0010 (JFB); The research foundation TSE-Partnership (JFB)
\item Authors' contributions: Conceptualization: MD, JFB, IR; Methodology: MD, JFB; Investigation: MD; Visualization: JF; Funding acquisition: IR; Project administration: MD; Supervision: IR; Writing – original draft: MD, JFB; Writing – review \& editing: MD, JFB, IR
\item Materials \& Correspondence: Correspondence and requests for materials should be addressed to M.D.
\item Additional information: Supplementary Information is available for this paper.

\end{itemize}

\bibliography{sn-bibliography}

\end{document}